\documentclass[aps,prl,twocolumn,preprintnumbers,groupedaddress]{revtex4}

\pdfoutput=1

\usepackage{todonotes}

\usepackage{amssymb}
\usepackage{epsfig} 
\usepackage{amsmath,color} 
\usepackage{graphicx} 
\usepackage{ulem} 
\usepackage{multirow}
\usepackage{tabularx}
\usepackage{slashed}

\usepackage{wrapfig}

\usepackage{tikz}
\usetikzlibrary{positioning,arrows}
\usetikzlibrary{decorations.pathmorphing}
\usetikzlibrary{decorations.markings}

\newcommand{\beq}{\begin{equation}}
\newcommand{\eeq}{\end{equation}}
\newcommand{\bea}{\begin{eqnarray}}
\newcommand{\eea}{\end{eqnarray}}
\newcommand{\ba}{\begin{array}}
\newcommand{\ea}{\end{array}}
\newcommand{\bi}{\begin{itemize}}
\newcommand{\ei}{\end{itemize}}
\newcommand{\bn}{\begin{enumerate}}
\newcommand{\en}{\end{enumerate}}
\newcommand{\bc}{\begin{center}}
\newcommand{\ec}{\end{center}}
\renewcommand{\l}{\left}
\renewcommand{\r}{\right}

\newcommand{\eq}[1]{Eq.~(\ref{#1})}
\newcommand{\eqs}[2]{Eqs.~(\ref{#1}) and (\ref{#2})}
\newcommand{\eqss}[3]{Eqs.~(\ref{#1}), (\ref{#2}) and (\ref{#3})}

\newcommand{\eV}{\mathinner{\mathrm{eV}}}

\newcommand{\MeV}{\mathinner{\mathrm{MeV}}}
\newcommand{\GeV}{\mathinner{\mathrm{GeV}}}

\begin{document}

\preprint{FTUV-17-08-13}
\preprint{IFIC/17-47}

\title{Lepton number asymmetries and the lower bound on the reheating temperature}


\author{Gabriela Barenboim$^1$}
\email[]{Gabriela.Barenboim@uv.es}

\author{Wan-Il Park$^2$}
\email[]{wipark@jbnu.ac.kr}
\affiliation{$^1$ Departament de F\'isica Te\`orica and IFIC, Universitat de Val\`encia-CSIC, E-46100, Burjassot, Spain}
\affiliation{$^2$ Division of Science Education and Institute of Fusion Science, Chonbuk National University, Jeonju 561-756, Korea}


\date{\today}

\begin{abstract}
We show that the reheating temperature of a matter-domination era in the early universe can be pushed down to the neutrino decoupling temperature at around $2 \MeV$ if the reheating takes place through non-hadronic decays of the dominant matter and neutrino-antineutrino asymmetries are still large enough, $|L| \gtrsim \mathcal{O}(10^{-2})$ (depending on the neutrino flavor) at the end of reheating. 
\end{abstract}


\maketitle


\section{Introduction}

Conventional wisdom assumes that the Universe was dominated by radiation soon after primordial inflation ended, until up the recent epoch of matter-radiation equality at an $ \eV$ scale background temperature.
However, in many well-motivated high energy theories, long-living particles are present, which can dominate the energy balance of the Universe for a while and decay at very late time.
Even the  inflaton, responsible for ending primordial inflation, can act as such a  particle. 
A matter-domination era arising due to the existence of a long-living non-relativistic particle is constrained by BigBang Nucleosynthesis (BBN) whose success requires the Universe to be dominated by a radiation composed of mostly standard model particles at 
its onset.
As a result, in principle, the reheating temperature (the temperature at the end of the last matter-domination era before BBN) could be pushed down to $1 \MeV$ scale.

Meanwhile, there have been a series of studies about the impact of a low reheating scenario on the properties of neutrinos in the early universe with the standard radiation background (see, for example, \cite{Kawasaki:2000en,Ichikawa:2005vw,deSalas:2015glj}). 
One of the key results in these studies is that the abundance of neutrinos is significantly suppressed as $T_{\rm R}$ is pushed down to $1 \MeV$ scale.
The reduction of the neutrino abundances around the epoch of BBN significantly affects the synthesis of light elements (especially the abundance of $^4He$), and is constrained by the cosmic microwave background (CMB) data \cite{Ade:2015xua} in terms of $N_{\rm eff}$, the number of relativistic degrees of freedom.
Hence the reheating temperature turns out to be more constrained than the naive expectation  and gets pushed to  $T_{\rm R} \gtrsim 5 \MeV$ \cite{deSalas:2015glj} \footnote{See also \cite{Choi:2017ncz} for a stronger bound coming from dark matter physics}.
This bound may look robust, since, even if some exotic extra radiation component may be introduced, making CMB constraint irrelevant, it would affect BBN in the wrong direction.
However, a sizable amount of lepton number asymmetries in neutrino background may relax the bound, since there are now two independent parameters $\Delta N_{\rm eff}$ (the extra radiation contribution from the asymmetries of three neutrino mass-eigenstates \footnote{The asymmetries are defined for mass-eigenstates although neutrinos due to their frequent interactions in the very early universe should be considered to be flavor-eigenstates. They won't become effectively incoherent mass eigenstates until after BBN time}) and $\xi_e$ (the degeneracy parameter of electron-neutrino) to play with in order to bypass both BBN and CMB constraints.

In this paper, we show that the reheating temperature can be pushed down to the neutrinos' decoupling temperature at around $2 \MeV$ if the neutrino-antineutrino asymmetries are large enough, where large means, $|L_{\alpha \alpha}| \gtrsim \mathcal{O}(10^{-2})$ ($\alpha = e, \mu, \tau$) at the end of reheating. 
This paper is organized as follows.
At first, the equations of motion for relevant fluid contents (including neutrinos) are presented and discussed.
Then, the results of the  numerical analysis are shown.
Finally, conclusions are drawn.

\section{Evolution equations in a low reheating scenario}

As the focus of our analysis is  a matter-domination era ending around the epoch of BBN, we consider the fluid contents to be made of a non-relativistic matter particle species (denoted as $\phi$), electromagnetic plasma ($\gamma, e^\pm, \mu^\pm$), and three neutrino/antineutrino flavors.
In this case, the expansion rate is given by
\beq
3 H^2 M_{\rm P}^2 = \rho_\phi + \rho_{\rm r}
\eeq
where $\rho_\phi$ is the energy density of $\phi$, and $\rho_{\rm r}$ is the radiation energy density associated to all the relativistic particles present.  
For $\phi$, the evolution equation is given by 
\beq
\dot{\rho_\phi} + 3 H \rho_\phi = - \Gamma_\phi \rho_\phi
\eeq
where the over-dot ` $\dot{}$ ' represents the time-derivative, and $\Gamma_\phi$ is the decay rate of $\phi$.
Following Refs.~\cite{Ichikawa:2005vw,deSalas:2015glj}, we assume that $\phi$ decays to light relativistic particles other than neutrinos without hadronic channels.

The thermal radiation background is characterized by the photon temperature $T_\gamma$.
As neutrino decoupling normally takes place around the epoch of BBN, it should be handled appropriately. In order to do that the safest way is to use the evolution equation of $T_\gamma$ deduced from the continuity equation of the  total energy density of the universe \cite{Ichikawa:2005vw,deSalas:2015glj}.
Numerically, this approach is a very precise way of tracking $T_\gamma$, provided that the evolution of the neutrino energy density is properly calculated at any time.
The subtlety here is that the decoupling of neutrinos depends on the momentum of the different modes and flavor mixing even complicates things further.
In summary, the presence of the energy injection from  the decay of the dominating matter particle together with the mode-dependent mixing and the need to track the neutrino decoupling leaves only the numerical approach as the proper way to attack the issue at hand. 
However, if all the modes decouple nearly simultaneously, the aforementioned complications can be avoided.
As will be discussed later, this phenomenon of simultaneous decoupling, is naturally expected when large neutrino-antineutrino asymmetries are present. 
In such a  case, the evolution of the radiation density $\rho_{\rm r}$ can be handled in the standard way which will be explained below.   

In the presence of an energy injection from the decay of $\phi$, the evolution equation of the radiation background is
\beq \label{eom-radiation}
\dot{\rho}_{\rm r} + 4 H \rho_{\rm r} = \Gamma_\phi \rho_\phi
\eeq
We define $T_{\rm i}$ and $T_{\rm R}$ to be the temperatures at the beginning and end of the matter-domination era, respectively.
At the early stage of matter domination, the fractional energy injection to the radiation density in the decays of $\phi$ particles is still small.
However, as time goes on, at a temperature $T_*$ the energy injection starts being the dominant source of the radiation density.
Hence,
$\rho_{\rm r}$ evolves as
\beq \label{rad-at-highT}
\rho_{\rm r} = \rho_{\rm r, i} \l(a_{\rm i}/a \r)^4
\eeq
for $T \gtrsim T_*$, and,
\beq \label{rad-from-decay}
\rho_{\rm r} = \frac{2 \sqrt{3}}{5} \Gamma_\phi M_{\rm P} \rho_\phi^{1/2}
\eeq   
for $T_{\rm R} \lesssim T \lesssim T_*$ where $T_*$ is given by
\beq \label{Tstar}
T_* = \l( \frac{g_*(T_{\rm R})}{g_*(T_*)} \r)^{1/4} \l( \frac{g_*(T_{\rm i})}{g_*(T_{\rm R})} \r)^{1/20} \l( \frac{T_{\rm i}}{T_{\rm R}} \r)^{1/5} T_{\rm R}
\eeq
Our definition of the reheating temperature $T_{\rm R}$ corresponds to
\beq
H (T_{\rm R}) \equiv \frac{\sqrt{2 \rho_{\rm r}(T_{\rm R})}}{\sqrt{3} M_{\rm P}} = \frac{2 \sqrt{2}}{5} \Gamma_\phi
\eeq
which gives
\beq
T_{\rm R} = \l( \frac{6}{5 \pi} \r)^{1/2} \l( \frac{10}{g_*(T_{\rm R})} \r)^{1/4} \sqrt{\Gamma_\phi M_{\rm P}}
\eeq
This definition of the reheating temperature differs from the more conventional one appearing in Ref.~\cite{deSalas:2015glj} where $\Gamma_\phi = 3 H(T_{\rm R})$ was used, but the resulting reheating temperature in our definition is larger only by a factor $\sqrt{6/5}\simeq 1.1$ which does not cause any sizable mismatch in comparison to the earlier works. 
If all the neutrinos were decoupled from the electromagnetic thermal plasma at temperatures $T_* < T < T_{\rm i}$, they would evolve independently  with negligible energy gain from the decay of $\phi$.
However, for $T_{\rm R} \sim \mathcal{O}(1) \MeV$ which is the focal point of our analysis, the decoupling is expected to take place for $T \lesssim T_*$ unless $T_{\rm i}$ is very close to $T_{\rm R}$.
We therefore assume that all neutrino modes decouple simultaneously from thermal bath at a temperature denoted as $T_{\rm dec}(<T_*)$.
This assumption will be justified later.
It is imporatnt to notice that after neutrino decoupling, $\rho_{\rm r}$ in \eq{rad-from-decay} contains $\gamma$ and $e^\pm$ exclusively.

In the early universe, the transformations of the different neutrino flavors can be described by the evolution of neutrino/anti-neutrino $3\times3$ density matrices $\rho$ and $\bar{\rho}$. 
For a mode of momentum $p$, when the universe is dominated by radiation, the evolution equations are given by \cite{Sigl:1992fn,Pantaleone:1992eq}
\bea 
\label{eom-rho}
i \frac{d\rho_p}{dt} &=& \l[ \Omega + \sqrt{2} G_F \l( \rho - \bar{\rho} \r), \rho_p \r] + C \l[ \rho_p \r] \\
\label{eom-rhobar}
i \frac{d \bar{\rho}_p}{dt} &=& \l[ -\Omega + \sqrt{2} G_F \l( \rho - \bar{\rho} \r), \rho_p \r] + C \l[ \bar{\rho}_p \r]
\eea  
In the above equations,
\beq \label{Omega}
\Omega = \frac{M^2}{2 p} - \frac{8 \sqrt{2} G_F p E_\ell}{3 m_W^2}
\eeq
where $M^2$ is the mass-square matrix of neutrinos in the flavor-basis, $G_F$ the Fermi constant, $m_W$ the mass of $W$-boson, $E_\ell = {\rm diag}(E_{ee}, E_{\mu\mu},0)$ the energy density of charged leptons, $\rho = (1/2 \pi^2) \int_0^\infty \rho_p p^2 dp$ (and similarly for $\bar{\rho}$), and $C[\dots]$ is the collision term.
For the mass-square matrix $M^2$, we consider normal hierarchy with masses and mixing angles for the neutrinos given by \cite{Agashe:2014kda,NOvA}
\bea
\Delta m_{21}^2 &=& 7.53 \times 10^{-5} \eV^2 
\\
\Delta m_{31}^2 &\simeq& \Delta m^2_{32} =  2.67 \times 10^{-3} \eV^2
\eea
and 
\beq
\theta_{12} = \frac{\pi}{5.5}, \ \theta_{23} = \frac{\pi}{4.6}, \ \theta_{13} = \frac{\pi}{20}
\eeq
where $\theta_{ij}$ are the mixing angles in the Pontecorvo-Maki-Nakagawa-Sakata (PMNS) mixing matrix \cite{Maki:1962mu,Pontecorvo:1957cp} whose CP-violating phase is set to zero \footnote{The effect of CP-violation is irrelevant in our discussion (see, for example, \cite{Barenboim:2016jxn}.)}.
Also, we take for simplicity $C[\rho_p] = - i D_{\alpha \beta} [\rho_p]_{\alpha \beta}$ for $\alpha \neq \beta$ only, and similarly for $C[\bar{\rho}_p]$ \cite{Dolgov:2002ab}.

In a matter-dominated universe, \eqs{eom-rho}{eom-rhobar} need to be modified.
At first, we consider the case of $T_{\rm R} \geq T_{\rm dec}$.
In this case, the temperature of the neutrinos is the same as the photon temperature until the epoch of $e^+$-$e^-$ annihilation at $T \sim 0.1 \MeV$.
So, in terms of temperature and momentum dependences, \eq{Omega} and the damping term can be used without modification for $T \gtrsim T_{\rm dec}$.
However, the self-interaction contained in \eqs{eom-rho}{eom-rhobar} is different from the one used in a radiation-dominated universe and such a modification has to be incorporated to our analysis.
Since $\rho-\bar{\rho}$ is simply given by $ \Delta n$, i.e., the difference in the number densities of neutrinos and antineutrinos, and there are no lepton number violation  interactions present at low energies, it should scale as $\l( \Delta n \r)_{\alpha \beta} \propto a^{-3}$ since the time its was generated at high enough energy. On the other hand, the photon number density behaves as $n_\gamma \propto T^3 \propto a^{-9/8}$ as matter-domination becomes close to its end. 
Hence, one finds
\bea \label{dn}
\l. \Delta n \r|_T &=& \l. \Delta n \r|_{T_i} \l( \frac{a_i}{a} \r)^3 
\nonumber \\
&\approx& \l\{
\begin{array}{ll}
\l. \Delta n \r|_{T_i} \l( \frac{T}{T_i} \r)^3 & {\rm for} \  T_* \leq T < T_i
\\
\l. \Delta n \r|_{T_*} \l( \frac{T}{T_*} \r)^8 & {\rm for} \ T_{\rm R} \leq T < T_*
\\
\l. \Delta n \r|_{T_{\rm R}} \l( \frac{T}{T_{\rm R}} \r)^3 & {\rm for} \  T < T_{\rm R}
\end{array}
\r.
\eea
where the $g_*(T)$ dependence has been ignored.

Well before the commencement of neutrino oscillations, each neutrino flavor is expected to be in thermal equilibrium, i.e., its distribution function in the massless limit amounts to
\beq
f(p,\mu,T)=\l(e^{\frac{p-\mu}{T}}+1 \r)^{-1}
\eeq
where $\mu$ is the chemical potential. 
Hence, in the flavor basis of $\l\{\nu_e, \nu_\mu, \nu_\tau \r\}$, when neutrino interactions with the  thermal background are still frequent as it is the case in the very early universe, the initial configuration of $\rho-\bar{\rho}$ can be expressed as 
\beq
\rho-\bar{\rho} = {\rm diag} \l( \Delta n_e, \Delta n_\mu, \Delta n_\tau \r)
\eeq
where 
\beq \label{dn-ini}
\Delta n_\alpha = \frac{\pi T^3}{3} \l[ \l( \frac{\xi_\alpha}{\pi} \r) + \l( \frac{\xi_\alpha}{\pi} \r)^3 \r] 
\eeq
with $\xi_\alpha \equiv \mu_\alpha/T$ being the degeneracy parameter of $\nu_\alpha$ for $T > T_i \gg \mathcal{O}(10) \MeV$.

\begin{figure}[h!]
\begin{center}
\includegraphics[width=0.49\textwidth]{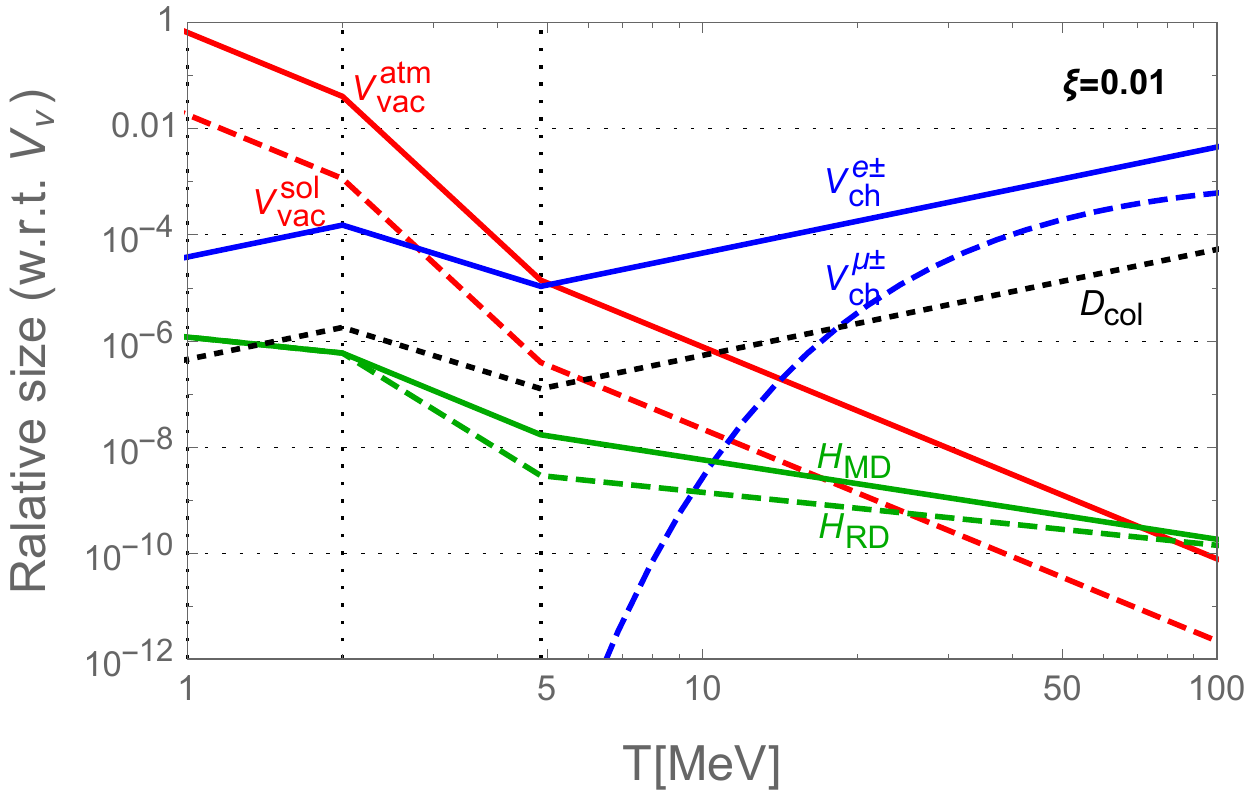}
\caption{Relative size of the potentials $V_{\rm vac}, V_{\ell^\pm}$, the momentum-averaged collisional damping $D_{\rm col}$, and the expansion rate $H$ with respect to the self-potential $V_\nu$ for $\xi=0.01$.
Matter-domination is assumed to start at a temperature $T_i = 150 \MeV$. 
The reheating temperature was set to be $T_{\rm R}=2 \MeV$.
`atm' and `sol' identify the contributions arising  from mass-square differences associated to atmospheric and solar neutrino oscillations respectively.
`MD' and `RD' stand respectively for cases of matter- and radiation-domination. 
For $D_{\rm col}$, we consider the terms associated to $\nu_{\mu,\tau}$  exclusively to stay on the most conservative side.
}
\label{fig:ratios}
\end{center}
\end{figure}

Now, we can evaluate the relative size of the different contributions driving  the neutrino evolution: the vacuum potential (sourced by the neutrino mixing), the one arising from the charged thermal background, the collision terms in \eqs{eom-rho}{eom-rhobar} for the averaged momentum mode, and the expansion rate to the neutrino self-interaction potential (i.e., the terms proportional to $\rho-\bar{\rho}$ in \eqs{eom-rho}{eom-rhobar}). Such a comparison is shown in Fig.~\ref{fig:ratios} where
\bea
V_{\rm vac}^{\rm atm} \equiv \Delta m_{\rm atm}^2 / \l( 2 \langle p \rangle \r), && V_{\rm vac}^{\rm sol} \equiv \Delta m_{\rm sol}^2 / \l( 2 \langle p \rangle \r)
\\
V_{\rm ch}^{e^\pm} \equiv \frac{8 \sqrt{2} G_F \langle p \rangle \l[ E_{\ell l} \r]_{11}}{3 m_W^2}, && V_{\rm ch}^{\mu^\pm} \equiv \frac{8 \sqrt{2} G_F \langle p \rangle \l[ E_{\ell l} \r]_{22}}{3 m_W^2} 
\\
V_\nu \equiv \sqrt{2} G_F \Delta n_\nu, && D_{\rm col} \equiv D_{e \mu}
\eea
with $\Delta n_\nu $ being a typical value of $\l( \Delta n \r)_{\alpha \beta}$ obtained from \eqs{dn}{dn-ini}.
From the  comparison of $H_{\rm MD, RD}$ and $D_{\rm col}$ in the figure, it is clear that the kinetic decoupling of a typical neutrino mode  is expected to take place at $T_{\rm dec} = 1 - 2 \MeV$ as long as $T_{\rm R} \gtrsim 2 \MeV$.
Therefore the situation is identical to that of a radiation-dominated universe.
At the same time, from the solid redline (left-top) in the figure, it is clear  that as long as $|\xi_\alpha| \gtrsim 10^{-2}$ at $T_i = 150 \MeV$, the self-interaction term always dominates over the vacuum potential of neutrinos at least until $T \gtrsim T_{\rm dec}$.
This implies that, if a lepton number asymmetry on the neutrino background  was generated, for example, during the first radiation-dominated era,  before the matter-domination epoch we are considering, for $T \gtrsim T_{\rm dec}$ the self-interaction potential would dictate the evolution of the neutrino energy density  as it contribution vastly dominates over the other contributions to the neutrino energy as long as 
\beq \label{xi-ini}
\l| L_\alpha^i \r| \approx \l| L_\alpha^f \r| \l( \frac{T_i}{T_{\rm R}} \r) \gtrsim 10^{-2} \l( \frac{T_i}{T_{\rm R}} \r)
\eeq
where $L_\alpha^i$ and $L_\alpha^f$ are the initial asymmetries and the ones at the neutrino decoupling, respectively.
Interestingly, it has been found that, if the self-interaction potential is large enough relative to the vacuum potential, i.e., 
\beq
\kappa \equiv \frac{V_\nu}{{\rm Max}\l[V_{\rm vac}^{\rm atm}, V_{\rm vac}^{\rm sol}\r]} \gg 1 
\eeq 
all neutrino modes are synchronized to the momentum-averaged mode and behave like a single compound system in its evolution which is basically dictated by neutrino flavor-mixings \cite{Samuel:1993uw,Pastor:2001iu,Abazajian:2002qx}.
Hence, conservatively speaking, for $\kappa \gtrsim \mathcal{O}(10)$ (i.e., $|L_\alpha| \gtrsim \mathcal{O}(10^{-2})$ around the epoch of decoupling), the synchronization will be a good enough approximation and we can safely ignore the distortions on the spectral distribution of neutrinos during the process of flavor-mixing and decoupling.
In what follows, we will consider this case for simplicity.
This approach of course is precise only at the percent level, but such an accuracy  is well enough to prove our point.

\section{Results of numerical integrations} 

\begin{figure*}[ht!]
\begin{center}
\includegraphics[width=0.48\textwidth]{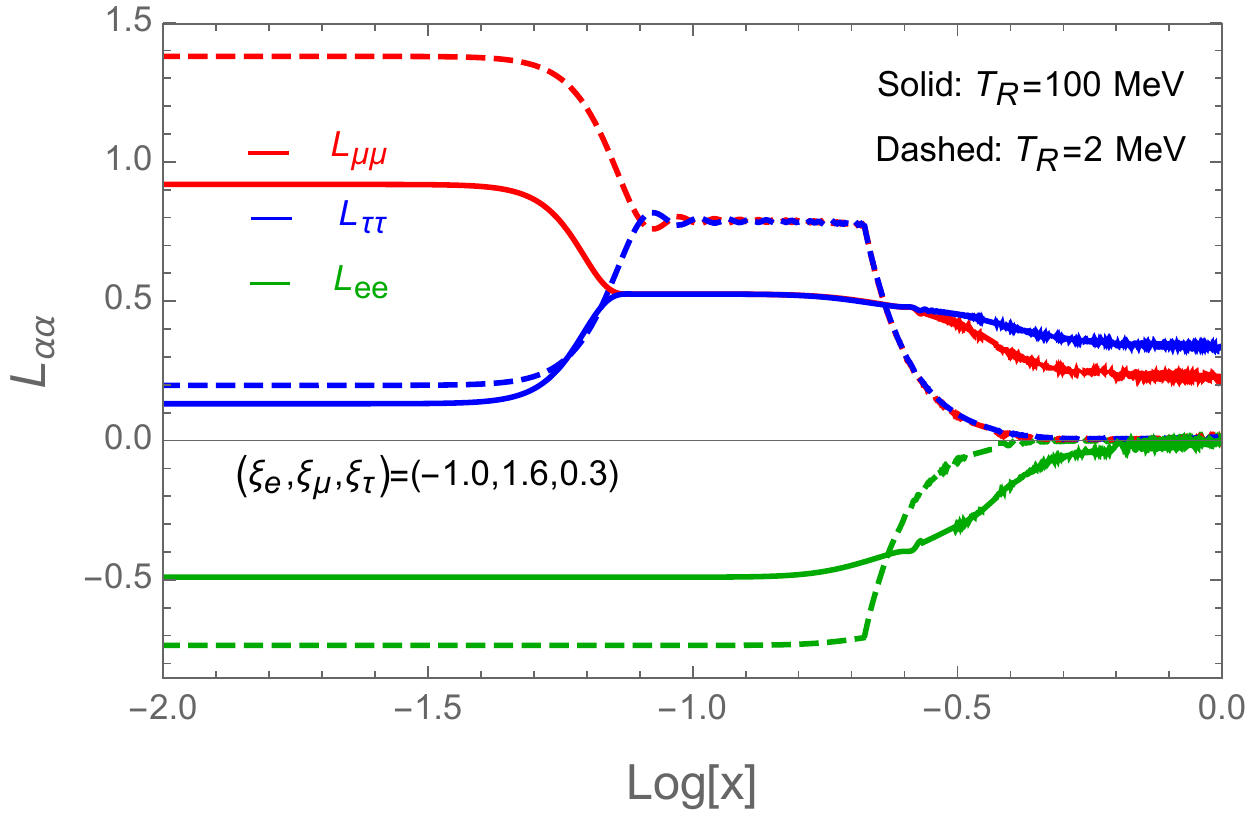}
\includegraphics[width=0.48\textwidth]{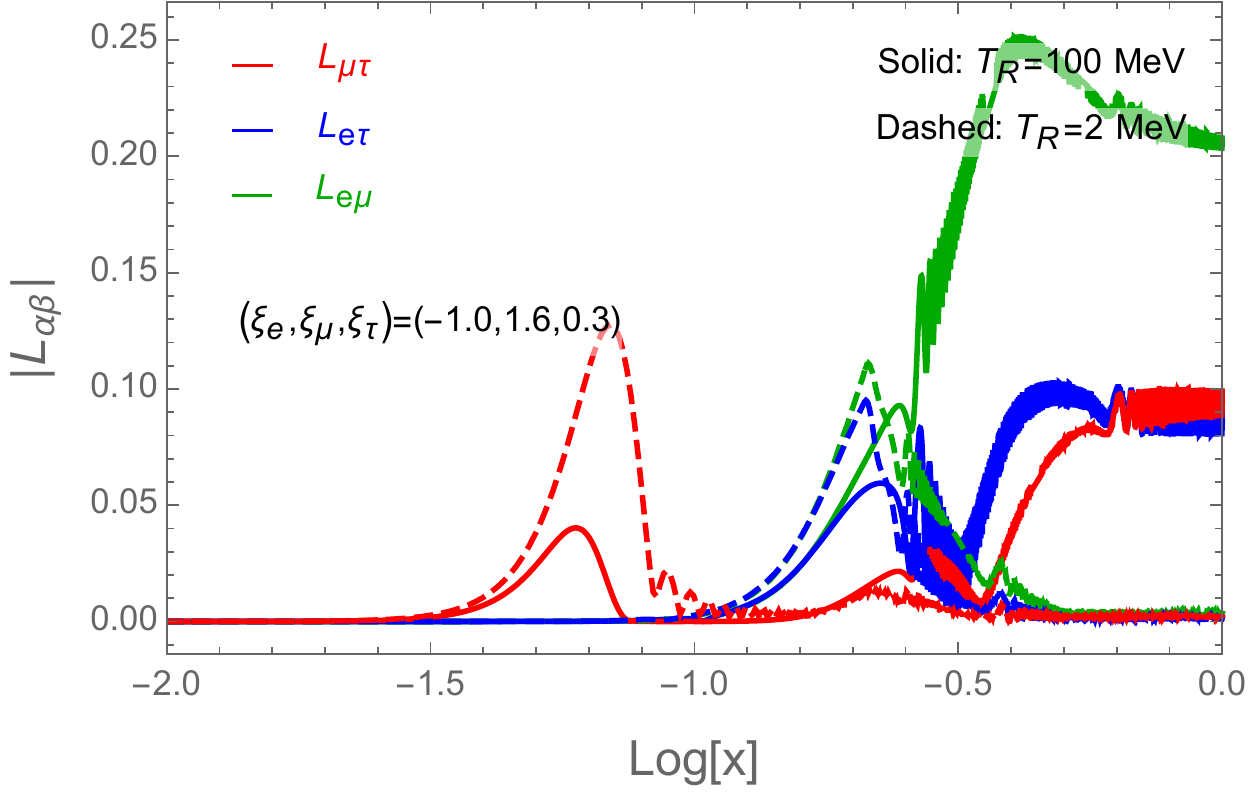}
\includegraphics[width=0.48\textwidth]{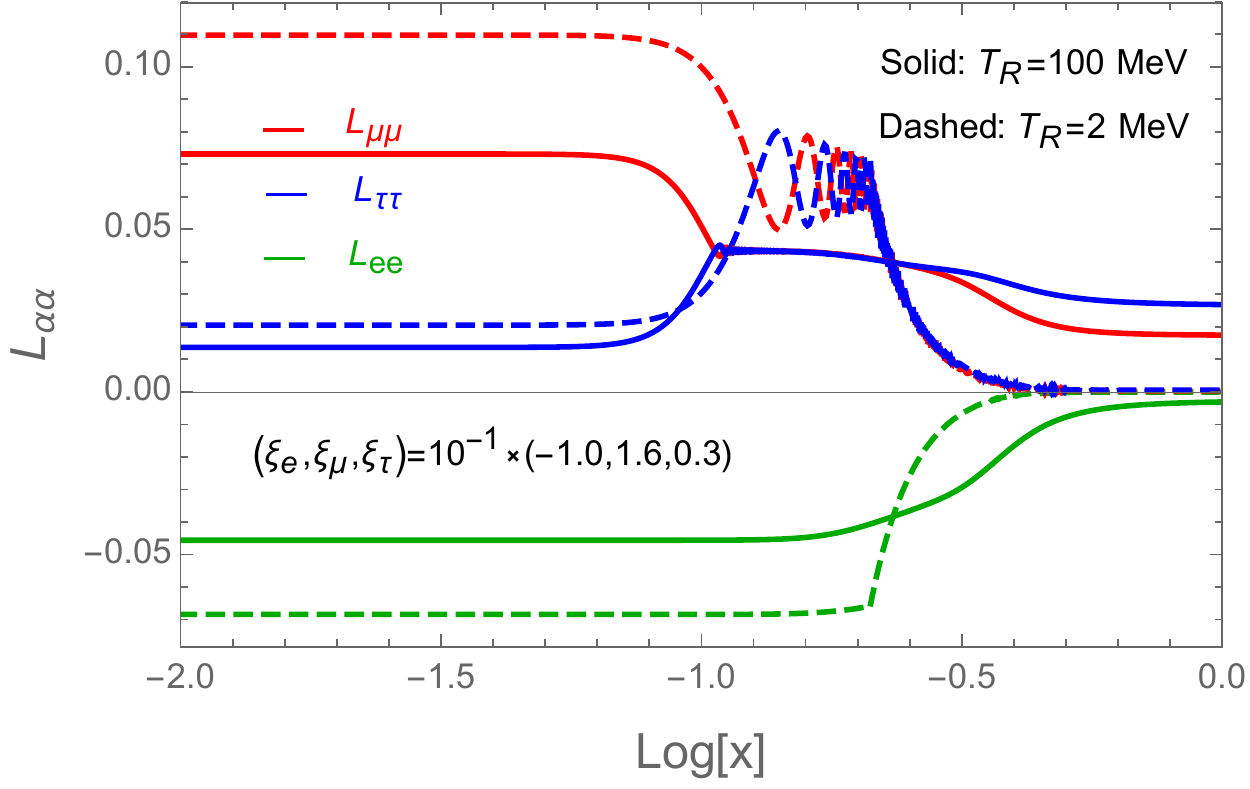}
\includegraphics[width=0.48\textwidth]{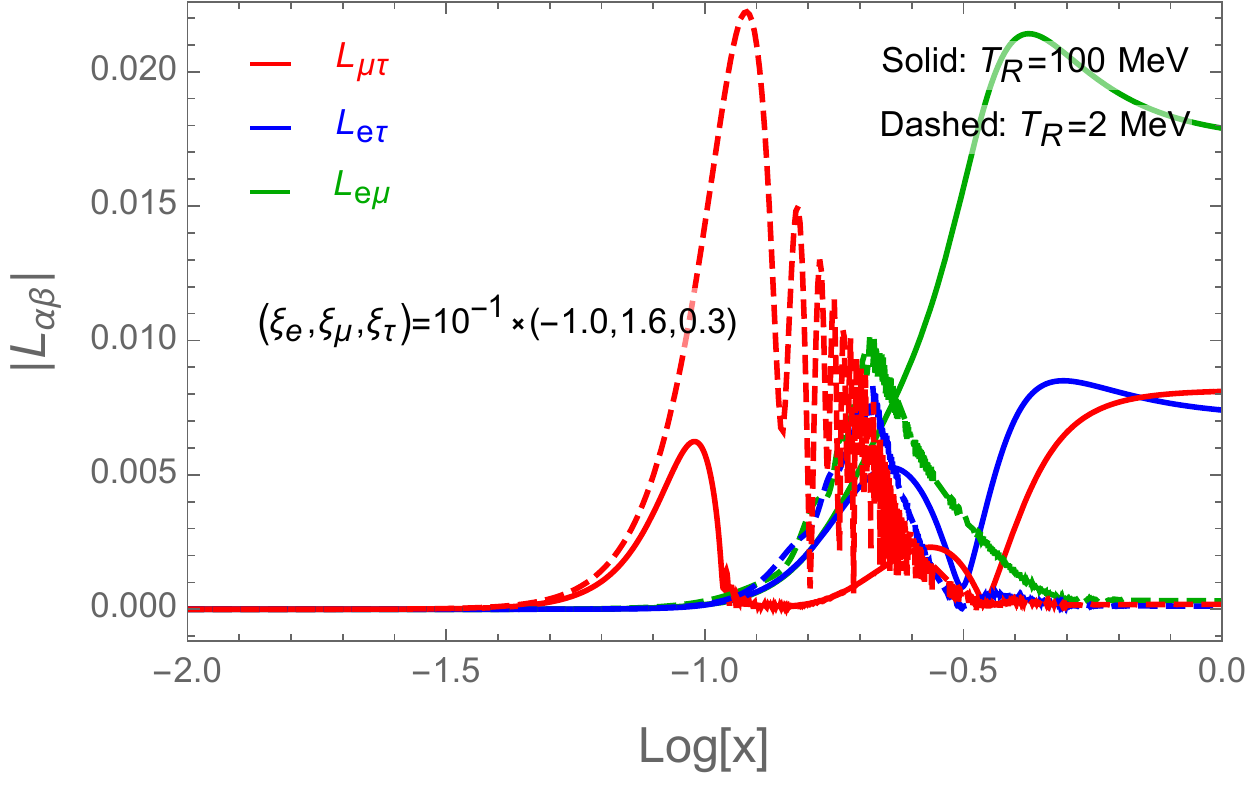}
\caption{Evolutions of lepton number asymmetries defined as $L \equiv \l( \rho - \bar{\rho} \r)/n_\gamma$.
\textit{Left}: Diagonal entries (pure flavor states). 
\textit{Right}: Off-diagonal entries (flavor-mixed states). 
Notice that off diagonal entries do not vanish proving that once neutrino interactions are no longer effective, neutrinos propagate as effectively incoherent mass eigenstates, {\it i.e.} the flavour basis is no longer a good description of the system.
}
\label{fig:Td-110-vs-2}
\end{center}
\end{figure*}

When the evolution equations of the neutrino density matrices (\eqs{eom-rho}{eom-rhobar}) is numerically computed, quite often one rescales the momentum mode $p$ by multiplying it by the scale factor $a$.
In a radiation-dominated background without any energy injection from other sources to the energy content, the rescaled momentum is a comoving momentum and therefore it is constant throughout the evolution of the universe.
However, in matter-dominated universe with an energy injection to radiation background present, the momentum of a neutrino is redshifted as $a^{-3/8}$ before its decoupling and as $a^{-1}$ after it.
Also, for large neutrino degeneracies, all modes are expected to be highly synchronized.
So, we can ignore the deviation of each neutrino flavor's distribution function from thermal one.
In this case, a convenient choice for the rescaled momentum is
\beq
y \equiv p/T_\nu
\eeq
which is constant during the evolution of the universe across the neutrino decoupling irrespective of the nature of the energy dominating the universe, matter or radiation.

In the numerical integration of \eqs{eom-rho}{eom-rhobar}, the initial condition was set as
\beq \label{rho-p-initial}
\rho_p = f(y,0)^{-1} {\rm diag}(f(y,\xi_e), f(y,\xi_\mu), f(y,\xi_\tau)), 
\eeq
and similarly for $\bar{\rho}_p$ but with $\xi_\alpha \to -\xi_\alpha$.
We then recasted the evolution equations in terms of the scale factor $a$ and took the single mode approach as in Ref.~\cite{Barenboim:2016shh}.
For the evolution of the photon temperature \eqss{rad-at-highT}{rad-from-decay}{Tstar} were used as an approximation in order to simplify an already complicated calculation.

The result of the integration is shown in Fig.~\ref{fig:Td-110-vs-2} where we compare the cases of $T_{\rm R} = 100 \MeV$ and $2 \MeV$ for $|\xi_\alpha| \sim 1$ and $0.1$ with $T_i = 150 \MeV$.
In the figure, one can see that, compared to the case of $T_{\rm R} = 100 \MeV$, a significant change on lepton number asymmetries $L_{\alpha \alpha}$ (diagonal entries) and $|L_{\alpha \beta}|$ (off-diagonal entries) for $T_{\rm R} = 2 \MeV$ appears.
Such drastic change is simply due to the significant energy injection (or entropy release) emanating from the decay of the dominating matter particle. 
Note that, as long as the neutrinos are decoupled after the reheating is completed, the symmetric components of neutrinos and antineutrinos will provide nearly the same amount of radiation energy to the thermal background as the standard case of radiation domination.
That is, it is expected to have 
\beq
N_{\rm eff, sym}^{\rm MD} \simeq N_{\rm eff, sym}^{\rm RD}
\eeq
for $T_{\rm R} \geq T_{\rm dec}$.
On the other hand, $\Delta N_{\rm eff}^{\rm MD}$ coming from lepton number asymmetries is obtained as follows with the flavor-mixing effects ignored for simplicity. 
For $1 \MeV \lesssim T \lesssim T_{\rm dec} < T_{\rm R}$ the symmetric component of the energy density of a neutrino flavor $\nu_\alpha$ is simply given by
\beq \label{rhonu-sym}
\frac{\rho_\alpha^{\rm MD}(T)}{\rho_\alpha^{\rm MD} (T_i)} = \l( \frac{T}{T_i} \r)^4
\eeq
However, since lepton number violating process is supposed to be shut down well before the matter domination, the number density of the asymmetric component of a neutrino species scales as $a^{-3}$ and the corresponding energy density is given by
\beq \label{rhonu-asym}
\frac{\Delta \rho_\alpha^{\rm MD}(T)}{\Delta \rho_\alpha^{\rm MD}(T_i)} =  \frac{T \Delta n_\alpha (T)}{T_i \Delta n_\alpha(T_i)} = \l( \frac{T}{T_i} \r) \l( \frac{a_i}{a} \r)^3
\eeq
Hence, from \eqs{rhonu-sym}{rhonu-asym} with
\beq
\l( \frac{a_i}{a} \r)^3 
= \l( \frac{T_*}{T_i} \r)^3 \l( \frac{T_{\rm R}}{T_*} \r)^8 \l( \frac{T}{T_{\rm R}} \r)^3
\eeq
and \eq{Tstar}, we find 
\beq \label{dNeff-highTR}
\Delta N_{\rm eff}^{\rm MD}(T) \equiv \sum_\alpha \l( \frac{\Delta \rho_\alpha}{ \rho_\alpha} \r)^{\rm MD}_T = \l. \Delta N_{\rm eff}^{\rm MD} \r|_{T_i} \l( \frac{T_{\rm R}}{T_i} \r)
\eeq
where the summation is for three neutrino flavors, and the change in the number of relativistic degrees of freedom for $1 \MeV \lesssim T < T_i$ was ignored.
Since $\l. \Delta N_{\rm eff}^{\rm MD} \r|_{T_i} = \l. \Delta N_{\rm eff}^{\rm RD} \r|_{T_i}$, we find
\beq
\Delta N_{\rm eff}^{\rm MD} \ll \Delta N_{\rm eff}^{\rm RD}
\eeq
Note also that, even if the constraint for a successful BBN may require $|L_{\alpha \alpha}| \lesssim 10^{-2}$ in a general sense, $|L_{\mu \mu, \tau \tau}| \gg \mathcal{O}(1)$ with $|L_{ee}| \lesssim 10^{-2}$ is perfectly allowed as discussed in Ref.~\cite{Barenboim:2016shh} and results in a larger  expansion rate at present time \footnote{The late time $\Delta N_{\rm eff}$ coming from neutrino asymmetries should not be calculated as the one associated to flavor-eigenstates but that of mass-eigenstates if only diagonal entries are going to be summed as explained in \cite{Barenboim:2016lxv}. Clearly $ \Delta N_{\rm eff}$   is independent of the  basis, but its trace  does not describe $ \Delta N_{\rm eff}$ correctly if written in the flavour basis 
as it does not account for the off-diagonal elements which arise due to neutrino 
oscillations and also contribute to the final asymmetry.}. 

If $T_{\rm R} < T_{\rm dec}$, \eqs{eom-rho}{eom-rhobar} should be handled carefully to take into account the difference between temperatures of neutrinos and photons. 
In the presence of very large asymmetries in the neutrino background  the decoupling of all modes takes place simultaneously, and its distribution can be considered thermal to an excellent approximation.
Once decoupled, the energy density of neutrinos is redshifted as $a^{-4}$ while that of photons is redshifted as $a^{-3/2}$ until $T \to T_{\rm R}$.
Hence, one finds that for the photon temperature in the range of $1 \MeV \lesssim T \lesssim T_{\rm R}$
\beq
\l( \frac{\rho_\alpha}{\rho_\gamma} \r)^{\rm MD}_T = \l( \frac{\rho_\alpha}{\rho_\gamma} \r)_{T_{\rm dec}}^{\rm MD} \l( \frac{a_{\rm dec}}{a_{\rm R}} \r)^{5/2}
\eeq
where
\bea
\rho_\alpha^{\rm MD}(T) &=& \rho_\alpha^{\rm MD}(T_{\rm dec}) \l( \frac{a_{\rm dec}}{a} \r)^4
\nonumber \\
&=& \rho_\alpha(T_i) \l( \frac{a_i}{a_*} \r)^4 \l( \frac{a_*}{a_{\rm dec}} \r)^{\frac{3}{2}} \l( \frac{a_{\rm dec}}{a} \r)^4
\eea
Also,
\bea \label{drho-nu-MD}
\frac{\Delta \rho_\alpha^{\rm MD}(T)}{\Delta \rho_\alpha(T_i)} 
&=& \frac{\Delta \rho_\alpha^{\rm MD}(T_{\rm dec})}{\Delta \rho_\alpha(T_i)} \l( \frac{a_{\rm dec}}{a} \r)^4
\nonumber \\
&=& \l( \frac{a_i}{a_*} \r)^4 \l( \frac{a_*}{a_{\rm dec}} \r)^{\frac{3}{8}}  \l( \frac{a_*}{a_{\rm dec}} \r)^3 \l( \frac{a_{\rm dec}}{a} \r)^4
\eea
As a result, the effective number of neutrinos is obtained as 
\bea \label{Neff-sym-lowTR}
N_{\rm eff, sym}^{\rm MD}(T)  
&=& \l. N_{\rm eff, sym}^{\rm RD} \r|_{T_{\rm dec}} \l( \frac{a_{\rm dec}}{a_{\rm R}} \r)^{5/2} 
\nonumber \\
&=& \l. N_{\rm eff, sym}^{\rm RD} \r|_{T_{\rm dec}} \l( \frac{T_{\rm R}}{T_{\rm dec}} \r)^{20/3}
\eea
and, defining $\Delta N_{\rm eff}^{\rm MD}(T) \equiv \sum_\alpha \Delta \rho_\alpha^{\rm MD}(T)/\rho_\alpha^{\rm RD}(T)$ for a comparison to the case of the conventional radiation domination, from \eqs{drho-nu-MD}{Tstar} one finds
\beq \label{dNeff-lowTR}
\Delta N_{\rm eff}^{\rm MD}(T) = \l. \Delta N_{\rm eff}^{\rm MD} \r|_{T_i} \l( \frac{T_{\rm R}}{T_i} \r) \l( \frac{T_{\rm R}}{T_{\rm dec}} \r)^{5/3}
\eeq
In \eq{dNeff-lowTR}, it was assumed that the energy density of asymmetric neutrinos is at most comparable to the other radiation density at $T=T_i$.
If it were the dominant component of radiation density, \eq{dNeff-lowTR} should be modified.
When the asymmetric neutrinos are subdominant radiation, $\l. \Delta N_{\rm eff}^{\rm MD} \r|_{T_i} = \l. \Delta N_{\rm eff}^{\rm MD} \r|_{T_*}$.
Also, when they dominate the radiation energy, $\l. \Delta N_{\rm eff}^{\rm MD} \r|_{T_*}$ is given in a simple form.   
Hence, a clearer formula for $\Delta N_{\rm eff}^{\rm MD}(T)$ applicable to both cases is found to be  
\beq \label{dNeff-lowTR-2}
\Delta N_{\rm eff}^{\rm MD}(T) = \l. \Delta N_{\rm eff}^{\rm MD} \r|_{T_*} \l( \frac{T_{\rm R}}{T_i} \r) \l( \frac{T_{\rm R}}{T_{\rm dec}} \r)^{5/3}
\eeq
where $\l. \Delta N_{\rm eff}^{\rm MD} \r|_{T_*}$ is given by
\bea \label{dNeff-Tstar}
\l. \Delta N_{\rm eff} \r|_{T_*} 
&=& \frac{15}{7} \sum_\alpha \l( \frac{\xi_{\alpha,*}}{\pi} \r)^2 \l[ 2 + \l( \frac{\xi_{\alpha,*}}{\pi} \r)^2 \r]
\\ \label{dNeff-Tstar-bnd}
&\leq& (4/7)\times g_*(T_*)
\eea
with $\xi_{\alpha,*} \equiv \xi_\alpha(T_*)$ and $g_*(T_*)$ being the degeneracy parameter of a neutrino species at $T_*$ and the number of relativistic degrees of freedom, respectively.
From \eqss{dNeff-lowTR-2}{dNeff-Tstar}{dNeff-Tstar-bnd}, if one requires the upper bound of $\Delta N_{\rm eff}^{\rm MD}(T)$ to be at least comparable to the would-be observed total effective number of neutrinos, $N_{\rm eff, tot}^{\rm obs} \sim 3$, the temperature at the onset of matter domination is upper-bounded as
\beq
\frac{T_i}{T_{\rm R}} \lesssim \frac{g_*(T_*)}{N_{\rm eff, tot}^{\rm obs}} \frac{4}{7} \l( \frac{T_{\rm R}}{T_{\rm dec}} \r)^{5/3} = \mathcal{O}(1-10)
\eeq    
where it was used that, even if $T_{\rm R}$ is close to $1 \MeV$, $T_{\rm dec} \lesssim 2 \MeV$ for a numerical estimation.
This result implies that, when $T_{\rm R} < T_{\rm dec}$, it is difficult for the asymmetric neutrinos to compensate the energy reduction of the symmetric component of neutrinos unless $T_{\rm R}$ is quite close to $T_i$ (i.e., the period of matter-domination is quite short so as for its effect to be negligible).
Therefore, practically (i.e., when $T_i \gg T_{\rm R}$) $T_{\rm R}$ is lower-bounded as 
\beq
T_{\rm R} \geq T_{\rm dec} \sim 2 \MeV
\eeq

\begin{figure}[t!]
\begin{center}
\includegraphics[width=0.49\textwidth]{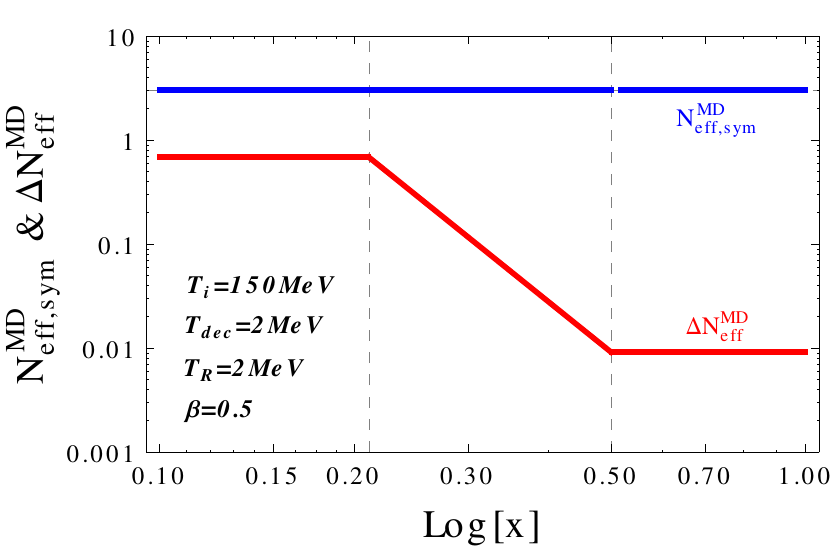}
\caption{$N_{\rm eff, sym}^{\rm MD}$ (blue line) and $\Delta N_{\rm eff}^{\rm MD}$ (red line) as functions of $x \equiv \MeV/T$ for $T_{\rm R} = T_{\rm dec} = 2 \MeV$.
The vertical dashed line correspond to $T_*$ and $T_{\rm dec}$ from left to right.
The effect of flavor-mixing was taken into account by introducing an additional suppression factor $\beta$ \cite{Barenboim:2016shh} in the right-hand side of \eqs{dNeff-highTR}{dNeff-lowTR}
}
\label{fig:Neff-vs-T}
\end{center}
\end{figure}
\begin{figure}[ht!]
\begin{center}
\includegraphics[width=0.48\textwidth]{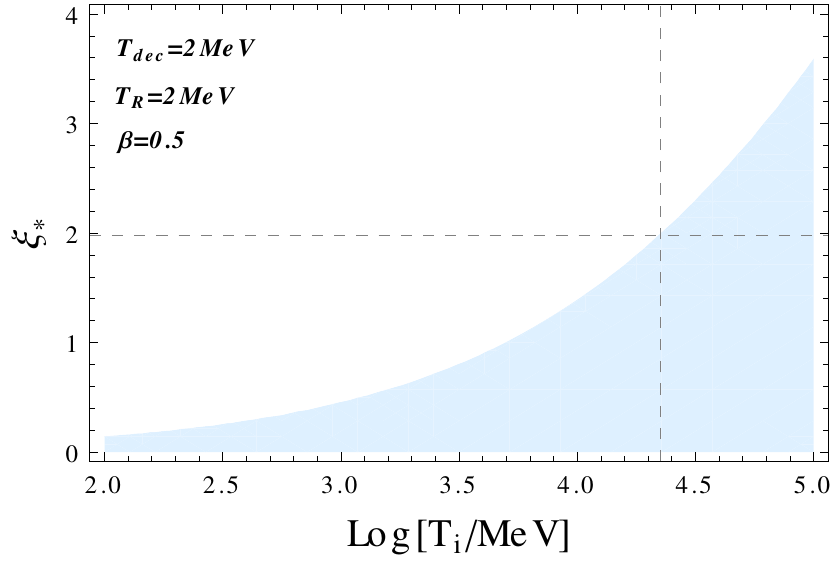}
\caption{
Bounds on the degeneracy parameter $\xi_{\alpha,*}$ of neutrinos as a function of $T_i$.
$|\xi_{\alpha,*}|=\xi_*$ was assumed for all neutrino flavors.
The upper boundary of the blue region corresponds to $|L_{\alpha\alpha}| = 10^{-2}$ at $T<T_{\rm R}$, chosen to guarantee the mode-synchronization due to strong self-interactions of neutrinos.
The horizontal dashed line corresponds to the possible maximal value of $\xi_*$ at $T=T_*$ with $g_*(T_*)=10.75$ for example. 
Hence, the allowed region of $\xi_*$ is the region below the horizontal dashed line but above the boundary of the blue region. 
}
\label{fig:xi-vs-TR}
\end{center}
\end{figure}
Fig.~\ref{fig:Neff-vs-T} shows $N_{\rm eff,sym}^{\rm MD}$ and $\Delta N_{\rm eff}^{\rm MD}$ as functions of $x\equiv \MeV/T$ for $T_{\rm R}= T_{\rm dec} = 2 \MeV$ as an example. 
As shown in the figure, even if the net contribution of neutrinos to radiation density is required to be similar to the case of the usual radiation domination, for $T_{\rm R} \geq T_{\rm dec}$ the asymmetric component of neutrinos does not need to be the major contribution to the neutrino energy density.
For the same set of $T_{\rm R}$ and $T_{\rm dec}$, in Fig.~\ref{fig:xi-vs-TR} we depicted the required $\xi_{\alpha,*}$ as a function of $T_i$.
From the figure, one can see that, if $T_{\rm R}$ is as low as $T_{\rm dec}$ and $|L_{\alpha\alpha}| \gtrsim \mathcal{O}(10^{-2})$ is required, $T_i$ is upper-bounded as
\beq
T_i \lesssim 20 \GeV
\eeq

It should be noted however that the simultaneous occurrence of extended periods of early matter domination and large asymmetries at reheating time are challenging, as primordial exponentially large asymmetries would need to be generated (however see Ref.~\cite{Barenboim:2017dfq} as an example of such a case).

\section{Impact on BBN}  
We have argued that, when $T_{\rm R} \ll T_i$ in a matter-dominated universe before BBN, $T_{\rm R}$ is lower-bounded at $T_{\rm dec}$ in order for the symmetric component of neutrino energy density not to be suppressed even though the asymmetric component gets through a large suppression.
Even if it can not be the main contribution to the energy density of neutrinos, large enough asymmetries of $|L_{\alpha \alpha}| \gtrsim \mathcal{O}(10^{-2})$ for $T \geq T_{\rm dec}$ can maintain the thermal distribution of the neutrino spectrum through  neutrino self interactions , allowing a right (or large enough) amount of radiation energy density for a successful BBN and a correct  matching to CMB data.

\section{Conclusions}

In this paper, we showed that the reheating temperature of a matter-domination era in the early universe can be safely pushed down close to the decoupling temperature of neutrinos $T_{\rm dec} \sim 2 \MeV$ if neutrino-antineutrino asymmetries normalized by the number density of photons are kept large enough, say $\mathcal{O}(10^{-2})$ until up to the neutrinos' decoupling. 
 
It has been known that, in the presence of a matter-domination era close to the BBN epoch, the standard contribution of neutrinos to the radiation background is reduced due to (i) the departure from thermal distribution which is induced by momentum dependent  early decoupling of neutrinos and (ii) the late time entropy release emerging from  the decay of the dominating matter particle.
This triggers two effects: (i) smaller expansion rate for a given photon temperature after reheating and (ii) smaller weak interaction rate caused by the reduced neutrino distribution function.
The latter seems dominant, causing the earlier decoupling of weak interactions and an increase of the Helium abundance ($Y_p$) as an unwanted consequence. 
Hence, $T_{\rm R}$ is constrained by BBN and CMB (from the constraint on $N_{\rm eff}$) to be $T_{\rm R} \gtrsim 5 \MeV$.
However, this result does not hold in the presence  of lepton number asymmetries sourced by neutrinos.
If neutrino asymmetries exist, {\it i.e.} if neutrino asymmetries were generated at a very high scale, this picture drastically changes.
The key aspect is that, if neutrino asymmetries are quite large such as $L_{\alpha \alpha} \gtrsim \mathcal{O}(10^{-2})$ until up to $T_{\rm dec} \sim 2 \MeV$, all the modes of the neutrino ensemble in the early universe get synchronized, behaving like a single compound system, and are expected to simultaneously decouple from the electromagnetic thermal bath. 
In this case the deviation of the neutrino spectrum from a thermal distribution is expected to be negligible.
BBN constraint can be still satisfied by keeping the asymmetry of electron-neutrinos small, although this fact would require some tuning of the initial configuration of the neutrino asymmetries.

\section{Acknowledgements}
GB acknowledges support from the MEC and FEDER (EC) Grants SEV-2014-0398, FIS2015-72245-EXP, FPA2014-54459, and the Generalitat Valenciana under grant PROMETEOII/2013/017. 
WIP acknowledges hospitality of Theoretical Physics Department at University of Valencia where a part of this work was performed.
This project has received funding from the European Union's Horizon 2020 research and innovation program under the Marie Sklodowska-Curie grant Elusives ITN agreement No 674896  and InvisiblesPlus RISE, agreement No 690575 and by research funds for newly appointed professors of Chonbuk National University in 2017, and by Basic Science Research Program through the National Research Foundation of Korea (NRF) funded by the Ministry of Education (No. 2017R1D1A1B06035959).


\end{document}